\documentclass{article}
\usepackage{spconf,amsmath,graphicx}
\usepackage{amsthm}
\usepackage{multirow}
\usepackage{booktabs}
\usepackage{adjustbox}
\usepackage{tabularx}
\usepackage{makecell}
\usepackage{cite}
\usepackage{amssymb}
\usepackage{stmaryrd}
\usepackage{ upgreek }
\usepackage[title]{appendix}
\usepackage{bm}

\usepackage[top=1in, left=0.75in, right=0.75in]{geometry}
\title{Fraudulent User Detection Via Behavioral Information Aggregation Network (BIAN) on Large-Scale Financial Social Network}
\name{Hanyi Hu, Long Zhang, Shuan Li, Zhi Liu, Yao Yang, Chongning Na\textsuperscript{*}}
\address{Fintech Research Center, Zhejiang Lab}
\begin{document}

%
\maketitle
\begin{abstract}
Financial frauds cause billions of losses annually and yet it lacks efficient approaches in detecting frauds considering user profile and their behaviors simultaneously in social network . A social network forms a graph structure whilst Graph neural networks (GNN), a promising research domain in Deep Learning, can seamlessly process non-Euclidean graph data .  In financial fraud detection, the modus operandi of criminals can be identified by analyzing user profile and their behaviors such as transaction, loaning etc. as well as their social connectivity. Currently, most GNNs are incapable of selecting important neighbors since the neighbors’ edge attributes (i.e., behaviors) are ignored. In this paper, we propose a novel behavior information aggregation network (BIAN) to combine the user behaviors with other user features. Different from its close “relatives” such as Graph Attention Networks (GAT) and Graph Transformer Networks (GTN), it aggregates neighbors based on neighboring edge attribute distribution, namely, user behaviors in financial social network. The experimental results on a real-world large-scale financial social network dataset, DGraph, show that BIAN obtains the 10.2\% gain in AUROC comparing with the State-Of-The-Art models. 
\end{abstract}
\begin{keywords}
Anomaly Detection, Risk Management, Graph Neural Networks, Representation Learning, Graph Attention
\end{keywords}
\section{Introduction}
\label{sec:intro}

\indent Financial frauds have become a major social problem in the past decades and cause billions of losses worldwide every year. Moreover, criminals have diversified their modus operandi via mobile phone, making counterfeit identities, or even anonymous bitcoin transaction. Demand of anomaly detection is driven by regulatory organizations and better credibility of financial enterprises and institutions. \newline
\indent The precaution and supervision of financial institutions are mostly rule-based \cite{butgereit2021anti} by supervising initial user profile or thresholding the amount or frequency of transaction in practice. Though most such systems may have high throughput, strictly follows certain regulatory rules and good interpretability for domain experts, it tends to generate high false positive rate and have low generalizability in detecting criminals at the very beginning especially with camouflage. \newline
\indent Deep learning-based approaches \cite{krizhevsky2017imagenet}\cite{sutskever2014sequence}\cite{chorowski2015attention} have shown its advantages in many general-purpose tasks such as image processing, natural language processing, speech recognition. Recently, AI-based method also tackles realistic social problems including prediction of protein-structure \cite{jumper2021highly}, material property \cite{xie2018crystal}, battery life expectancy\cite{yang2021machine}. Graph neural networks (GNN)\cite{kipf2016semi}\cite{hamilton2017inductive}\cite{velivckovic2017graph} flourish since 2016 and been widely adopted in risk management\cite{yu2021abnormal}. \newline
\indent In financial fraud detection task, frauds are always the minority, leading to a highly imbalanced and partially labeled sample distribution. GNN\cite{kipf2016semi} was originally designed and specialized for semi-supervised tasks. GNN eases the problem of hazardous data problem to some extent by information propagated by neighbors. However, criminals tend to be adversarial and disguise their profile. It drives us utilizing other information such as user behaviors.To our best knowledge, we are the first work learn causality or relativeness between edges, aggregate edge representation in line with node attribute via ETNConv layer and use it for relevant node selection and weighting.
\indent In financial fraud detection task, frauds are always the minority, leading to a highly imbalanced and partially labeled sample distribution. GNN\cite{kipf2016semi} was originally designed and specialized for semi-supervised tasks. The essence of GNN \cite{kipf2016semi} is propagating information of neighboring nodes and learning target nodes representation by compensating missing information from those in proximity. It eases the problem of fraudulent user detection to some extent, especially for the criminal groups with subtle social connections. However, user profile could be disguised, as criminals tend to be adversarial. However, the behaviors of users can reflect their purpose and intention to some extent. Hence, we intend to better utilize behavior information such as action, temporal information on edges for weighting relevant neighboring nodes for the downstream node classification task. Most current GNNs focus on nodes’ attributes with edge information left unused or rather implicit by propagating node and edge information back and forth \cite{liu2022graph}, in this paper, we propose to weight neighboring nodes with similar behaviors. 

Our main contributions are summarized as follow:
\begin{itemize}
\item We proposed a framework, BIAN, which boost node aggregation process by learning inter-edge causality/relativeness induced edge representation and used a novel edge-to-node convolutional (ETNConv) layer to aggregate edge representation in line with target nodes.
\item We provided the theoretic proof of association between combining TGAT \& Fourier series encoding scheme and cross-correlation of two time series signals.
\item We have conducted extensive experiments on a real large-scale financial social networks dataset, DGraph, and outperformed other state-of-the-Art (SOTA) models by up to 10.2\% measuring in AUROC. 
\end{itemize}
The result of the paper will be unfolded as follows. In Section 2, we introduce other relevant works in the literature. In Section 3, we provide an overview of our proposed method. In Section 4, We will elaborate the experiment details on the real-world financial social network dataset, DGraph. Lastly, we conclude our work and state the possible future directions in section 5.\\

\section{Related Works}\label{sec:format}
	\textbf{Graph-based anomaly detection.} Many works \cite{chaican}\cite{dou2020enhancing} have proposed graph-based solution on anomaly detection problem, since the graph data is ubiquitous and GCNs [15, 16] have shown its superiority in many tasks. As GNNs are not originally designed for anomaly detection, researchers extended and redesigned their models from vanilla GCN. AMNet\cite{chaican} designed low- and high-frequency kernels in filtering anomalous signal and learn the importance of two sets of kernels adaptively. CARE-GNN\cite{dou2020enhancing} developed a neighboring node selector module based on reinforcement learning and propagate camouflaged fraudsters information selectively. However, their works consider node attribute as input only.\newline

	\textbf{Temporal representation learning.} There are three regimes of temporal representation learning. The first category is by temporal information conditioned random walk sampler \cite{lee2019temporal} and learn node representation in the Skip-Gram \cite{mikolov2013distributed} fashion. The second category is based on signal spectrum encoding. i.e. Fourier series expansion. Time2vec \cite{kazemi2019time2vec} decomposes time signal into a set of frequency basis. The last class is learning graph-based representation incorporating signal spectrum encoding \cite{xu2020inductive}. Specifically, authors focused on time-span for invariant temporal representation with respect to specific timestamp \cite{sankar2018dynamic}. Our work has been greatly inspired by their work and we provided theoretical analysis of its association to cross-correlation. \newline

	\textbf{Edge-To-Node aggregation.} We then researched several edge-to-node aggregation methods. Hypergraph Conv \cite{bai2021hypergraph} is a graph convolutional layer, which propagate nodes’ feature to hyperedges and subsequently propagate from hyperedge to nodes back. Similarly, a dual message passing mechanism \cite{liu2022graph} was proposed by learning node-to-edge and edge-to-node representation alternatively. However, they used edge only as an intermediate vector without considering edge-to-edge relevance.\newline

	\textbf{Edge-enhanced GCNs.}  The most natural way of incorporating edge information in GCNs is by using it as weights on adjacency matrix. EGNN \cite{gong2019exploiting} developed attention-based and convolution-based EGNN layer upon doubly stochastic normalized edge representation, which normalizes edge attributes feature-wise and node-wise. The attentive score is the product of the raw edge feature and attended node feature. Similarly, ECC \cite{simonovsky2017dynamic} and GTEA \cite{li2021temporal} have incorporated edge feature as attention score in a similar way, but on dynamic graph. ECC used a multi-layer perceptron (MLP) \cite{taud2018multilayer} or attention in edge feature aggregation, while GTEA proposed to use sequential model or a novel sparsity-induced attention in the edge aggregation step. The structure of our work is very similar to GTEA, however, we used a novel Edge-To-Node convolutional layer in edge aggregation stage and our edge aggregation are not just between edges with its source and target nodes, but also adjacent edges with at least one common connected node. 

\section{Method}
\label{sec:pagestyle}
\subsection{Preliminaries}
\label{ssec:subhead}
\indent In this work, we are mainly focusing on edge attributed graph. Edges are attributed by timestamp, edge type, etc. Formally, we can define the edge attributed graph as $G=(V, E, X_V, T_E)$. Specifically, $V=\{v_i \mid 0 \leqslant i \leqslant n\}$ and $E \in \{ e_j=(u,v) : \ 0\leqslant j \leqslant m \& u,v\in V\}$. The adjacency matrix A is a $n \times n$ matrix where $a_{ij}=1$ if $e_{ij} \in E$ and 0 otherwise. Each node $v_i$ are associated with $i_{th}$- row of node attribute $X_{V} \in R^{n \times d_v}$ where n is the node cardinality and $d_v$ is the node attribute dimension. Likewise, each edge $e_j$ are associated with $j_{th}$- row of edge attribute $T_E \in R^{m \times d_e}$ where m is the edge cardinality and $d_e$ is the edge attribute dimension. Temporal edge attributed graph $G_T$ is when timestamp is included in the edge attribute.  \newline
Incidence matrix $H \in \{0,1\}^{n \times m}$ can be analogous to adjacency matrix A, where $h_{ij} = 1$ if there exists a pairwise relation between $i_{th}$ node and $j_{th}$ edge. 

\begin{equation}
h(v,e)=\left\{
\begin{array}{cl}
1 \, &  if \quad v \in e, \\
0 \, &  if \quad v \notin e. \\
\end{array} \right.
\end{equation}

\subsection{Overall Structure}
\label{ssec:subhead}
\indent The overall structure of our proposed model consists of 3 components: neighboring edge representation learning, edge-to-node convolutional layer and fusion of edge representation. Figure 1 illustrates the overall structure of our model.

\begin{figure}[htb]
\centering
  \centerline{\includegraphics[width=8.5cm]{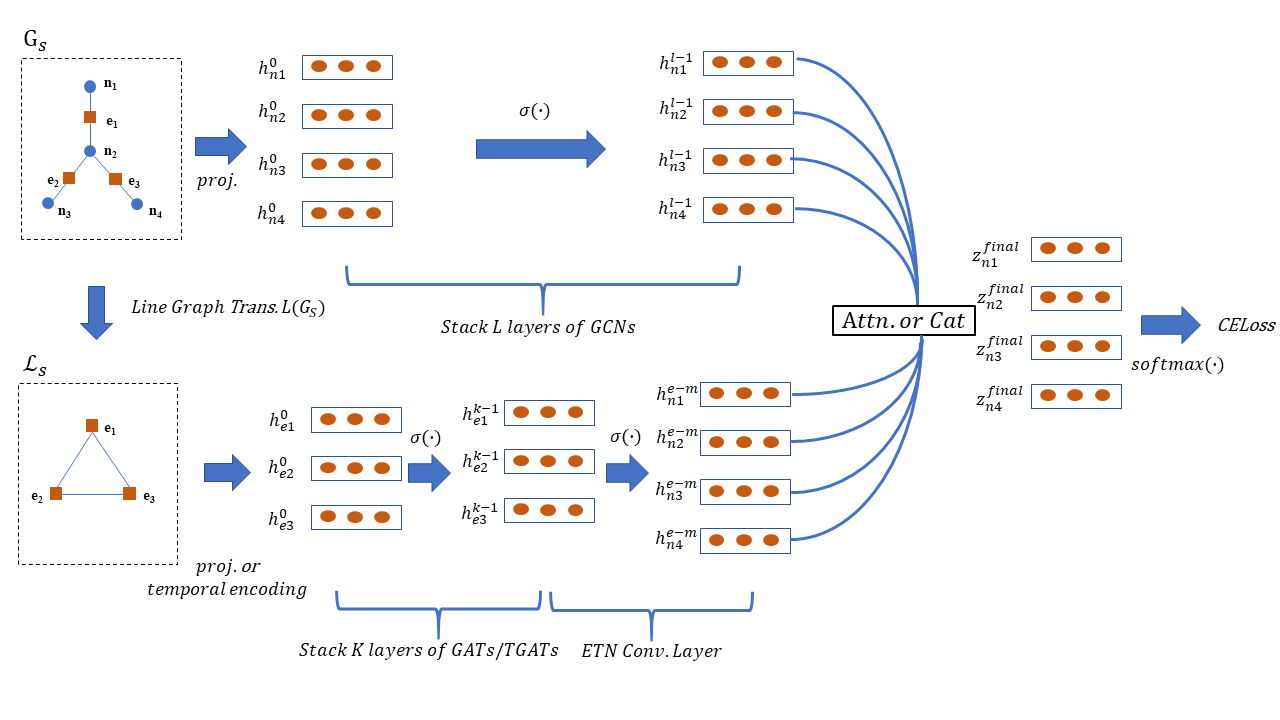}}
  \centerline{Fig1.Overall Structure of Our Proposed Model BIAN. }\medskip
\end{figure}
\vspace{-0.8cm}

\subsection{Neighboring Edge Representation Learning}
\label{ssec:subhead}
\indent Neighboring edge representation learning module is aiming for capturing most important edge information surrounding target nodes. We followed the neighboring sampling scheme\cite{hamilton2017inductive} and train GCNs on mini-batched subgraphs to avoid exceeding memory limitation on large graph. In light of learning relative edge importance, we convert mini-batched subgraph $\mathcal{G}_s$ to edge-dominated line graph representation ${L_{\mathcal{G}}}_s=\mathcal{L}(\mathcal{G}_S)$ to contruct an edge-oriented data representation to learn better inter-edge information. Line graph can be denoted as in eq (2).
\begin{equation}
  \begin{aligned}
  \mathcal{L}(\mathcal{G}) &= (\mathcal{V}',\mathcal{E}'), \, \mathcal{V}' = \mathcal{E}', \, \mathcal{E}'= \{(e_1,e_2) : e_1 \cap e_2\ \neq \emptyset\} \\
  \end{aligned}
  \end{equation}
Next, we aim to learn edge representation via inter-edge relevance. 
\subsubsection{Non-Temporal Edge Attribute Learning}\label{sssec:subsubhead}
We adopted the graph attention network (GAT)\cite{velivckovic2017graph} in encoding edge representation and the formulation of GAT is shown in eq(3),

 \begin{equation}
  \begin{aligned}
  	&{h_i'}^{(l+1)} = \sigma(\sum_{j\in N_i}^n\beta_{ij}'W_{v}{h_{j}'}^{(l)}),\beta_{ij}' = \frac{\exp{(e_{ij}')}}{\sum_{k\in N_k}^n\exp{(e_{ik}')}}  \\
	& e_{ij}'=\frac{Attn(QK^T)}{\sqrt{d}}, Q = W_q{h_i'}^{(l)}  , \, K = W_k{h_j'}^{(l)}
  \end{aligned}
  \end{equation}

where edge attention coefficients $e_{ij}'$ denoting the importance of source node $v_j'$ with respect to target node $v_i'$. $Attn$ is a shared attentional mechanism which maps from $R^{F'}\times R^{F'}$ to a comparable attention score $\beta_{ij}'\in R$. $W_q\in R^{d\times F'}$, $W_k\in R^{d\times F'}$ and $W_v\in R^{F'\times F'}$  are learnable parameters.

\subsubsection{Temporal Edge Attribute Learning}\label{sssec:subsubhead}
\indent For temporal edge attribute learning, we followed the work from \cite{sankar2018dynamic} which combined TGAT and Time2vec temporal encoding. Analogously, TGAT is formulated in eq (4) with an extra $M_{t_{i}'t_{j}'}$ term.
\begin{equation}
  \begin{aligned}
  	&{h_i'}^{(l+1)} = \sigma(\sum_{j\in N_i}^n\beta_{ij}'W_{v}{h_{j}'}^{(l)}),\beta_{ij}'=\frac{\exp{(e_{ij}' + M_{t_{i}'t_{j}'})}}{\sum_{k\in N_k}^n\exp{(e_{ik}  + M_{t_{i}'t_{k}'})}}  \\
	& e_{ij}'=\frac{Attn(QK^T)}{\sqrt{d}}, Q = W_q\varphi(t_i') , \,K =W_k\varphi(t_j')
  \end{aligned}
  \end{equation}
The temporal encoding scheme\cite{mikolov2013distributed}\cite{xu2020inductive} is defined as a series of sinusoidal basis:
\begin{equation}
  \begin{aligned}
t\shortrightarrow\varphi_{d}(t) := \sqrt{\frac{1}{d}}[\cos(w_1t), \sin(w_1t), ..., \cos(w_dt), \sin(w_dt)]
  \end{aligned}
  \end{equation}
\indent In Appendix A, we illustrated detailed proof of the association of cross-correlation with non-orthogonal frequency basis evaluated at $\uptau=t_i'-t_j'$ with edge attention coefficients $e_{ij}'$. The cross-correlation function measures the similarity of two signals as the displacement of one with respect to the other and can reflect the causality of two behaviors in the time domain to some extent. In Appendix A, we illustrated detailed proof. 
\subsubsection{Time-Conditioned Edge Attribute Learning}\label{sssec:subsubhead}
\indent We then explored time-conditioned attribute learning, where we kept TGAT as the encoder with temporal order constraint term $M_{t_{i}'t_{j}'}$, but uses projected edge attribute to calculate attention score $\beta_{ij}'$. Intentionally, the model can learn the edge relative importance under temporal order constraint.\newline
\indent We can apply different GCN variants to further encode multi-hop edge representation and get edge representation $Z_e$.
\subsection{ETNConv Layer}
\label{ssec:subhead}
\indent Aiming to complete the fraudulent node classification, we propose a novel Edge-To-Node convolutional layer to propagate message from edge to node and obtain the edge-aggregated node representation $Z_{e-merged}$.
The ETNConv Layer is defined as follows, \newline
\begin{equation}
  \begin{aligned}
	Z_{e-merged} = \sigma(D_n^{-1}HZ_eW)
  \end{aligned}
  \end{equation}
$Z_{e-merged}\in R^{n\times d_1}$ is the convolved edge-aggregated node representation, $\sigma$ denotes activation function, $H\in [{0,1}]^{(n\times m)}$ is the incidence matrix, $Z_e\in R^{m\times d_2}$ is the neighboring edge representation from previous section, $W\in R^{d_2\times d_1}$ is the learnable parameters and $D_n$ is the degree matrix of incidence matrix H. 
\begin{equation}
  \begin{aligned}
	D_n = \sum_{\upepsilon=1}^MH_{i\upepsilon}
  \end{aligned}
  \end{equation}
\subsection{Fusion of Edge Representation}
\label{ssec:subhead}
\indent Lastly, we need to merge original node attribute and edge-aggregated node representation. We could simply concatenate these two sets of representation or merge them by attention. 
The concatenated embedding is defined as, \newline
\begin{equation}
  \begin{aligned}
	Z_{final} = Concat(Z_{e-merged}, Z_n)
  \end{aligned}
  \end{equation}
\indent Or merge them by self-attention if one of the embedding is more dominant than the other. \newline
\begin{equation}
  \begin{aligned}
	&Z_{final} = Attn(Z_{e-merged}, Z_n) \quad Or \\ 
	&Z_{final} = Attn(Z_n, Z_{e-merged})  \\
  \end{aligned}
  \end{equation}

where,
\begin{equation}
  \begin{aligned}
	Attn(Q, K) = Softmax(\frac{QK^T}{\sqrt{d}})K
  \end{aligned}
  \end{equation}

\section{Experiments}
\label{sec:typestyle}
\indent In this section, we introduce the real-world dataset, DGraph, for elaborating the effectiveness and efficacy of the proposed model BIAN. Since we are making use of edge information and the relativeness between edges, we need attributes and timestamp associated with edge, and graph large in scale and authentic. The DGraph is the only compatible and publicly available dataset fulfills our requirement. We aim to answer the following research questions: \\
1) RQ1: How does the model perform comparing with other SOTA models in detecting fraudulent users?\\
2) RQ2: How does each module contributes to the performance gain?

\subsection{Dataset}
DGraph dataset\cite{huang2022dgraph} is a large scale financial social network dataset published by Finvolution Group with 3,700,550 nodes and 4,300,999 directed edges. There are 4 types of nodes, normal, anomaly and 2 types of background nodes. Anomaly nodes are users on their platform who overdue at least once, normal nodes are users who always pay on time and background nodes are those who registered and never borrow. The graph is highly sparse with average degree of 2.16 and they specifically introduced 2 types of background nodes to boost the connectivity between normal and abnormal nodes. The dataset is also highly imbalanced with anomaly ratio at 1.26\%. There are 17 desensitized node attributes based on user id, basic user profile, telephone number and emergency contacts. Each edge is associated with timestamp in 821 days window and belonging to 1 out of 11 types exclusively.\\

\subsection{Experiment Setup}
\label{ssec:subhead}
\indent We compared our proposed model with 4 classes of SOTA models on DGraph. 
\textbf{\emph{Baseline-Methods}} include MLP\cite{taud2018multilayer}, GCN\cite{kipf2016semi}, GraphSage\cite{hamilton2017inductive}, GraphSage-NeighSampler\cite{hamilton2017inductive}, GAT-NeighSampler\cite{velivckovic2017graph}, GATv2-NeighSampler\cite{brody2021attentive}, SIGN\cite{rossi2020sign}. Baseline models are mainly models that utilize nodes intrinsic features and its neighbors only. \newline
\textbf{\emph{GCN based Anomaly Detection Methods}}. AMNet\cite{chaican} is specifically designed for anomaly detection task, which learn predesigned low- and high- frequency kernels adaptively. However, their frequency kernels are applied on node attributes only.\newline
\textbf{\emph{Temporal GCN Methods}.} TGN\cite{kazemi2019time2vec} pretrains node attributes on dynamic graph and learn relative node importance in time domain. Again, it does not consider edge information. The TGN on the leader board reimplemented without pretraining process and replacing temporal convolution layer with graph transformer layer\cite{shi2020masked} named as TGN-no-mem in Table 1.   \newline
\textbf{\emph{Edge-enhanced Methods}.} GEARSage\cite{gearsage} is the best model on the leaderboard, which feature engineered extra node attributes and also encoded edge timestamp, edge direction, edge type. Nevertheless, they just concatenated edge feature to node and made them propagate by GraphSAGE layer altogether.\newline
\textbf{\emph{Implementation Details}}: All our experiments are implemented based on PyG and PyTorch package and trained on a machine with single 32G Tesla V100S GPU. All GNNs models are kept in 2 layers. Neighbor sampler from PyG samples from 2-hop neighboring nodes with 10 maximum neighbors in each layer. Accordingly, we set the number of GAT layer in node and edge branch to 2. We used Glorot initialization and Adam Optimizer for all models. All our results are repeated at least 3 times and report AUROC on testset.
\subsection{Result}
\label{ssec:subhead}
\subsubsection{Comparison with SOTA Models on DGraph (RQ1)}

\begin{table}[htbp]
 \centering
\begin{adjustbox}{width=\columnwidth,center}
    \begin{tabular}{lllllp{3cm}<{\centering}}
    \Xhline{1.5 pt}
    \multirow{2}[4]{*}{Model} & \multicolumn{4}{c}{Experimented Test AUROC } & \multicolumn{1}{c}{\multirow{2}[4]{*}{Reported Test AUROC}} \\
         & hidden size@16 & hidden size@32 & hidden size@64 & hidden size@128 &  \\
    \Xhline{1.5 pt}
    MLP\cite{taud2018multilayer}   & 71.29±0.2327  & 71.42±0.0504  & 71.73±0.1043  & 71.93±0.0743  & 71.92±0.0009 \\
    
    GCN\cite{kipf2016semi}   & 69.36±1.4500  & 71.39±0.2321  & 72.86±0.1499  & 73.13±0.2519  & 70.78±0.0023 \\
  
    GraphSAGE\cite{hamilton2017inductive}  & 73.53±0.3403  & 75.19±0.7625  & 76.23±0.0511  & 76.19±0.5383  & 77.61±0.0018 \\

    GraphSAGE-NeighSampler\cite{hamilton2017inductive} & 76.82±0.1010  & 77.47±0.0886  & 77.72±0.0586  & 77.75±0.0715  & 77.61±0.0018 \\

    GAT-NeighSampler\cite{velivckovic2017graph} & 73.54±0.1619  & 73.36±0.3609  & 73.39±0.0880  & 73.40±0.3020  & 73.33±0.0024 \\
    GATv2-NeighSampler\cite{brody2021attentive} & 76.43±0.7575  & 75.79±0.6748  & 76.47±0.1501  & 76.01±0.7947  & 76.24±0.0081 \\
    SIGN\cite{rossi2020sign}  & -     & -     & -     & -     & 78.23±0.0017 \\
    \hline
	AMNet\cite{chaican} & 73.80±0.0008  & 73.89±0.0008  & 74.05±0.0004  & 74.02±0.0004  & - \\
    \hline
    TGN-no-mem\cite{kazemi2019time2vec} & 79.66±0.0011  & 79.71±0.0006  & 79.81±0.0013  & x     & 77.41±0.0003 \\
    \hline
    GEARSage\cite{gearsage} & -     & -     & -     & -     & 84.60±0.0002 \\
    \hline
    BIAN with Timestamp & 94.76±0.0011  & 94.72±0.0006  & \textbf{94.79±0.0009}  & 94.62±0.0001  & - \\
    BIAN with Edge Attr. & 94.53±0.0016  & \textbf{94.55±0.0013}  & 94.50±0.0006  & 94.15±0.0004  & - \\
    BIAN with Time-Condi. Edge Attr. & 94.35±0.0002   & 94.45±0.0002  & 94.51±0.0002  & 93.89±0.0066  & - \\
    \hline
    \end{tabular}%
  \end{adjustbox}
  \label{tab:addlabel}%
\caption{Model Comparison on DGraph Dataset}
\end{table}%
\indent First, we found the result from baseline models together with AMNet are all around 75\%, which use node attribute only. The highest AUC is from GATv2-NeighborSampler at 76.47\%. Next, we examined the results from TGN-no-mem (with temporal information included) and GEARSage(with both temporal and edge type information included) with +3.34\% and +8.13\% improvement comparing with that of GATv2-NeighborSampler. It justified that temporal and edge-type features contribute to this particular task. We tested timestamp and edge attribute separately and obtained up to 18.3\% and 10.2\% improvement with respect to GATv2-NeighborSampler and GEARSage, which show the superioty of our model in learning these features.\newline
\indent We tested BIAN using timestamp, edge type and time-conditioned edge type encoding, which all had close performance. Therefore, our proposed model has shown level of generalizability on different types of edge attributes. From our experiments, we observed little fluctuation under different model size and small standard deviation in all settings, which verified the robustness of our model.

\subsubsection{Ablation Study (RQ2)}
\begin{table}[htbp]
  \centering
	\begin{adjustbox}{width=\columnwidth,center}
    \begin{tabular}{lllll}
    \Xhline{1.5 pt}
    \multirow{2}[4]{*}{Model} & \multicolumn{4}{c}{Experimented Test AUROC } \\
          & hidden size@16 & hidden size@32 & hidden size@64 & hidden size@128 \\
    \Xhline{1.5 pt}
    Edge Attr. Only & 52.69±0.0261  & 53.07±0.0301  & 51.77±0.0494  & 49.92±0.0145  \\
    Node Attr Only(sage) & 73.53±0.3403  & 75.19±0.7625  & 76.23±0.0511  & 76.19±0.5383  \\
    w/o Fusion by Attention & 76.12±0.0003  & 76.33±0.0020  & 76.22±0.0022  & 76.24±0.0014  \\
    w/o Edge Aggr. & 92.99±0.0004     & 92.59±0.0008     & \textbf{93.00±0.0007}     & 90.34±0.0328 \\
    BIAN by Random Freq. & 93.23±0.0160  & 91.74±0.0171  & 93.12±0.0149  & \textbf{94.38±0.0020}  \\
    BIAN by t2v & 94.76±0.0011  & 94.72±0.0006  & \textbf{94.79±0.0009}  & 94.62±0.0001  \\
    \Xhline{1.5 pt}
    \end{tabular}%
  \end{adjustbox}
  \label{tab:addlabel}%
	\caption{Ablation Study of BIAN}
\end{table}%
We then conducted an ablation study to evaluate the efficacy of each module. Firstly, we found similar performance under different frequecy initialization. As $W_Q$ and $W_K$ are learnable parameters, they can converge to the optimal setting by training. \newline
\indent Next, we removed second layer of GAT/TGAT in the neighboring edge representation stage. We observed a -1.79\% drop. Since the receptive field in line graph is larger than that in original graph. Hence, higher-order connectivity produced marginal gain in this case.\newline
\indent We then compared attention and concatenation in the fusion step, where there is a 18.46\% gap. In this sense, edge-aggregated node representation $Z_{e-merged}$ can boost neighboring node selection with similar behavior. The temporal or edge representation could potentially adjust the importance of neighboring nodes and guide message propagation when the graph is high in non-homophily. \newline

\section{Conclusion}
\label{sec:majhead}
\indent In this work, we proposed a behavior information aggregation network, BIAN, for fraudulent user detection in financial social network. BIAN learn embedding based on node and edge attribute encoded in separate branch. We structured the edge branch capable of learning the relativeness between edges by using GAT/TGAT on edge dominated line graph to incorporate causality between user behaviors in social network. Furthermore, we proposed a novel ETNConv Layer to aggregate learnt edge information aligning with node. We evaluated our method on a real-world large-scale financial social network, DGraph and outperformed current SOTA methods. The complexity in calculating edge feature is proportional to $E^2$. We will explore better way of calculating relative edge representation in the future. 

\begin{appendices}

\section{Appendix}
\textit{Lemma: The unnormalized attention score $\widetilde{e}_{ij}= QK^T$ in TGAT with temporal encoding is equivalent to cross-correlation with non-orthogonal frequency basis evaluated at $\uptau=t_i-t_j$.}
\begin{proof}
	The cross-correlation formula in discrete form is shown in eq(11),
	\begin{equation}
  		\begin{aligned}
			R({\uptau}) = R(t_i - t_j) = \sum_{m=-\infty}^\infty x[m]y[m+\uptau]
  		\end{aligned}
  	\end{equation}
	Since it is only a timestamp on the edge, indicating a Dirac function with respect to timestamp t, eq (1) is non-zero when $m=t_j$ and 0 elsewhere.
	\begin{equation}
	  	\begin{aligned}
			R(t_i-t_j)=x[t_i]y[t_j]
		\end{aligned}
	  \end{equation}
	Then, we can expand it by Fourier series, we can obtain eq (13).
	\begin{equation}
	  	\begin{aligned}
			R(t_i-t_j)=& [\sum_{n=0}^{d}a_n \cos(w_n t_i)+b_n \cos(w_n t_i)] \\
					+& [\sum_{n=0}^{d}c_n \cos(w_n t_i)+d_n \cos(w_n t_i)]
		\end{aligned}
	  \end{equation}
	given $W_Q:=[a_1,b_1,...,a_n,b_n]$ and $W_K:=[c_1,d_1,...,c_n,d_n]$. As $W_Q$ and $W_K$ are learnable parameters and the frequency projection weights $W_f:=[w_1,...,w_n]$ from the temporal encoding scheme can be optimized to some non-orthogonal sets.
\end{proof}
\end{appendices}

\bibliographystyle{IEEEbib}
\bibliography{reference}

\end{document}